\documentstyle[11pt]{article}
\newcommand{\nc}[1]{\newcommand{#1}}
\newcommand{\beq}{\begin{equation}}
\newcommand{\eeq}{\end{equation}}
\newcommand{\bea}{\begin{eqnarray}}
\newcommand{\eea}{\end{eqnarray}}

\def\gtwid{\raise.3ex\hbox{$>$\kern-.75em\lower1ex\hbox{$\sim$}}}
\def\ltwid{\raise.3ex\hbox{$<$\kern-.75em\lower1ex\hbox{$\sim$}}}
\nc{\gh}{\mbox{gh}}
\nc{\ngh}{\mbox{ngh}}

\begin{document}
\begin{titlepage}
\title{\Large{
{}From  Hamiltonian to  Lagrangian Sp(2) BRST Quantization}}
\vspace{0.5cm}
\author{{\sc K. Bering} \\
Institute of Theoretical Physics, Uppsala University,
S-751 08 Uppsala, Sweden
}

\maketitle
\begin{abstract}
We give a formal proof of the equivalence of Hamiltonian
and Lagrangian  BRST quantization.
This is done for a generic $Sp(2)$-symmetric theory
using flat (Darboux) coordinates.
A new quantum master equation is derived in a Hamiltonian setting
which contains all the Hamiltonian fields and momenta of a given theory.
\end{abstract}
\vfill
\begin{flushright}
UUITP-18/95
\\hep-th/9511020
\end{flushright}

\end{titlepage}
\newpage

\section{Introduction}
Both the Hamiltonian BRST quantization scheme \cite{BFV,h126}
and the Batalin-Vilkovisky Lagrangian BRST quantization scheme \cite{BV}
are well-established quantizations procedures.
It is therefore of major interest to give a general recipe for
the transition from a Hamiltonian BRST quantized system
to the corresponding BV Lagrangian BRST quantized system.
Many articles have been devoted to this problem  \cite{hehala}.
Grigoryan, Grigoryan and Tyutin extended \cite{ggt}
the phase space of a generic Hamiltonian theory
with some auxiliary fields and momenta,
introducing antifields as sources for the BRST transformation.
By integrating out the momenta, they were able to
formally derive a Lagrangian quantum master equation.
De Jonghe noted \cite{fdj} that the extension of the phase space
could be understood and organized in terms of collective fields,
using the prescription of Hamiltonian quantization.
In this manner the Schwinger-Dyson BRST symmetry is automatically
incorporated as first pointed out as a general principle by \cite{AD2}.
We improve this standard\footnote{The term {\em standard} is
merely used to distinguish from the $Sp(2)$ symmetric case,
where there also is an anti-BRST symmetry.}  BRST construction
further in section \ref{stdbrst}-\ref{stdqme}.
Here we show that, even in the enlarged Hamiltonian phase space
{\em without} integrating over the momenta,
a new quantum master equation is fulfilled.
Another improvement (of a bit more technical nature)
is that part of the kinetic terms ({\em 'momenta times velocity'})
in the effective action are in fact BRST-exact,
and can be removed.
It is remarkable that this little observation happens to
eliminate the necessity of the usual scaling procedure,
where the fields and momenta are redefined
by a power of a scaling parameter,
which in the end is scaled to infinity
to remove (some of the kinetic) terms \cite{h126,ggt,fdj}.

In the recent years $Sp(2)$-symmetric quantization schemes
have been formulated both in the
Hamiltonian case \cite{bltha2} and in the Lagrangian case \cite{bltla2,bm,bms}.
This is the main subject of this article.
One of the reasons why we treat the standard BRST case is to
show that many features are in fact already
present in the standard BRST case,
and can be directly transfered to the $Sp(2)$ case.
In section \ref{sp2brst}-\ref{sp2qme} we perform the transition from
Hamiltonian to Lagrangian picture for a general $Sp(2)$ theory.
The guiding principle is again the Schwinger-Dyson shift
symmetry \cite{DD,pkf}.
A new quantum master equation is derived in an extended phase space,
and the $Sp(2)$ Lagrangian theory of Batalin, Lavrov and Tuytin \cite{bltla2}
is formally extracted.

\section{Standard BRST}
\label{stdbrst}

In the case of standard BRST let us perform \cite{fdj}
the transition from a Hamiltonian system
to the equivalent BV Lagrangian system.
In order to achieve this we extend the phase space by introducing several new
fields,
especially those who plays the role of antifields in the Lagrangian picture.
Let us consider a Hamiltonian system with fields $\phi^A$ and momenta $\pi_A$,
a real\footnote{ The demand of Hamiltonian and BRST charge to be real
replaces the requirement of hermicity in a operator formulation.
In general, complex conjugation of supernumbers
corresponds to Hermitian conjugation of operators.
We emphasize that \mbox{$\overline{zw}=\overline{w} \ \overline{z}$}
for two supernumbers $z$, $w$.
And \mbox{$\overline{{\partial^l w} \over {\partial z}}
={ {\partial^r \overline{w}} \over {\partial \overline{z}}}$} if $w$ is a
function of $z$.}
Hamiltonian $H_o$ and a real BRST charge $\Omega_o$
with (right) BRST transformation
\mbox{$\delta_o =\left\{\cdot,\Omega_o\right\}$},
where \mbox{$\left\{\cdot,\cdot\right\}$} denotes the equal time Poisson
superbracket.
The Hamiltonian $H_o$ and the charge $\Omega_o$ fulfill
\bea
    \left\{\Omega_o(t),\Omega_o(t) \right\} &=&0 \cr
    \left\{H_o(t),\Omega_o(t) \right\}     &=&0 \cr
    {\partial \over \partial t}\Omega_o &=&0~.
\eea
The last equation, which states that there is no explicit time dependence
on the BRST charge $\Omega_o$, is a further requirement
that we impose to make the calculations simpler.
This is a rather weak assumption which is fulfilled for most theories of
interest.
The compact De Witt notation is used throughout this text,
i.e. summation and integration over repeated indices is implicitly understood.
An exception is in the above formula where the time-dependence
is explicitly mentioned in parenthesises. Then there is no integration over
time.

Note that $(\phi,\pi)$  runs over {\em all} fields of the given theory,
including the Lagrange multipliers, ghosts, and corresponding momenta.
We stress that the Hamiltonian  $H_o$ is the {\em full} BRST quantized
Hamiltonian.
There is a ambiguity in $H_o$ because one could always change
$H_o$ by a BRST exact amount.
If the theory has a classical Lagrangian formulation
with classical action $S_{\rm cl}$ (which does not depend on ghosts)
one could fix this ambiguity by demanding
\beq
     e^{{i \over \hbar} S_{\rm cl}}
=  \int \! {\cal D} \pi   \   e^{{i \over \hbar}
 \left(  \pi_A \dot {\phi}^A -  \int dt \  H_o \right)} ~.
\eeq
\label{classicallimit}
Even though this looks like a rather reasonable boundary condition to impose,
it is fairly complicated to study this boundary condition in a general setting.
In practice it may be that one cannot perform the $\pi$-integrations
explicitly,
or those turn out to be on the form
\beq
    \int \! {\cal D} \pi   \   e^{{i \over \hbar}
 \left(  \pi_A \dot {\phi}^A -  \int dt \  H_o \right)}
\ = \  {\rm delta~functions} \cdot e^{ {i \over \hbar} \   {\textstyle\rm
action}}  ~.
\eeq
A possible cure for the last problem is to reorganize
(fields$\leftrightarrow$momenta).
In general we have not done this analysis
and we give no guarantee that the boundary condition (\ref{classicallimit}) can
be met.
Fortunately, (\ref{classicallimit}) is not essential to the constructions given
below
other than to impose a classical limit.

The BRST transformation rules of the
phase space variables are organized as
\bea
    \delta_o \phi^A &=& i^{\epsilon_A} {\cal R}^A (\phi,\pi)  \cr
    \delta_o \pi_A &=&  {\cal R}_A (\phi,\pi) ~.
\eea
The factor $i$ is chosen such that the BRST {\em field} structure function
${\cal R}^A$ and the BRST {\em momenta} structure function ${\cal R}_A$ are
real.
%They share a rather similar notation, but should not be confused.
We further assume that  Hamiltonian theory is free of anomalies
in the sense that the Ward identities\footnote{It is enough to impose
the Ward identities for zero gauge fermion. In fact the principle that
BRST invariant Green functions do not depend on
the gauge fermion can be {\em derived} from this.}
\mbox{$< \delta_o G >_o=0$} is fulfilled
for every function \mbox{$G=G(\phi,\pi)$}.
At least this is very reasonable requirement seen
from an operator formulation point of view.
This leads (by a variational argument) to an equation
\beq
 \left(  i^{\epsilon_A}  { {\delta^r} \over {\delta\phi^A}}  {\cal R}^A
  +{ {\delta^r} \over {\delta\pi_A}} {\cal R}_A  \right)
   e^{{i \over \hbar}   \left(  \pi_A \dot {\phi}^A -  \int dt \  H_o \right)}
= 0 ~,
\label{hme}
\eeq
which can be viewed as a Hamiltonian master equation for the original theory.
It is assumed that all constraints $G_{\alpha}(\phi,\pi)$ are first class:
\bea
     \left\{G_{\alpha}(t),G_{\beta}(t) \right\}  &=&
C^{\gamma}_{\alpha\beta}(t) G_{\gamma} (t) \cr
     \left\{G_{\alpha}(t),H_o(t) \right\}  &=&  V^{\beta}_{\alpha}(t)
G_{\beta}(t)~.
\eea
We want to impose the Schwinger-Dyson shift symmetry \cite{AD2}.
Therefore, the phase space is extended with shift fields $\varphi^A_i$,
shift momenta $\pi^i_A$, shift ghosts ${\cal C}^A_i$ and shift antighosts
${\cal P}^i_A$.
The index $i=1,2$ is a sector index.
$i=1$ is often called the minimal sector,
while $i=2$ is referred to as the non-minimal sector.
Thus for each original field there are four new fields:
\beq
\begin{array}{lrcl}
    {\rm Fields} &\Phi^{\cal A} &=& \left\{ \phi^A, \varphi^A_i, {\cal C}^A_i
\right\} \\
    {\rm Momenta} &\Pi_{\cal A} &=& \left\{ \pi_A, \pi^i_A, {\cal P}^i_A
\right\} ~.
\end{array}
\eeq
We use the convention
\bea
     \overline{\Phi^{\cal A}}  &=& \Phi^{\cal A} \cr
     \overline{\Pi_{\cal A}}    &=&  (-1)^{\epsilon_{\cal A}} \Pi_{\cal A} ~,
\eea
so that Grassmann odd momenta is imaginary.
The Grassmann parities and ghost numbers are given by
\beq
\begin{array}{rcccccccl}
     \epsilon(\pi^i_A)&=&\epsilon(\varphi^A_i)&=&\epsilon(\pi_A)
     &=&\epsilon(\phi^A)&\equiv&\epsilon_A \\
     \epsilon({\cal P}^i_A)&=&\epsilon({\cal C}^A_i)&=&\epsilon_A+1 \\
    \\
     \gh(\varphi^A_i)&=&\gh(\phi^A)&\equiv&\gh_A \\
     \gh(\pi^i_A)&=&\gh(\pi_A)&=&-\gh_A \\
    \gh({\cal C}^A_i)&=&\gh_A+1&=& -\gh({\cal P}^i_A)  ~.
\end{array}
\eeq
The extended equal time Poisson superbracket reads:
\beq
\left\{ F(t),G(t) \right\} =
{ {\delta^r F(t)} \over {\delta\Phi^{\cal A}(t)} }
{ {\delta^l G(t)} \over {\delta\Pi_{\cal A}(t)} }
-(-1)^{\epsilon_{\cal A}}{ {\delta^r F(t)} \over {\delta\Pi_{\cal A}(t)} }
{ {\delta^l G(t)} \over {\delta\Phi^{\cal A}(t)} }~,
\eeq
or equivalently in terms of fundamental fields:
\bea
    \left\{ \Phi^{\cal A}(t),\Pi_{\cal B}(t) \right\}  &=&   \delta^{\cal
A}_{\cal B} \cr
     \left\{ \Phi^{\cal A}(t),\Phi^{\cal B}(t) \right\}  &=&  0 \cr
    \left\{ \Pi_{\cal A}(t),\Pi_{\cal B}(t) \right\}  &=&   0~.
\eea
As additional shift constraints $\chi^i_A$ we take
\bea
     \chi^1_A &=&  \pi_A + \pi^1_A  \cr
     \chi^2_A  &=&  \pi^2_A ~.
\label{chipi}
\eea
For an arbitrary function $F$ of the extended phase space variables
 $(\Phi^{\cal A} ,\Pi_{\cal A} )$ a shift operation $\sim$ is defined
\cite{fdj} as
\beq
F(\phi,\pi,\varphi_1,\pi^1,\varphi_2,\pi^2,\cdots)
\stackrel{\sim}{\longrightarrow}
\tilde{F}\equiv F(\phi-\varphi_1,\pi-\chi^1,0,0,\cdots)
=F(\phi-\varphi_1,-\pi^1,0,0,\cdots)~.
\eeq
This operation has several nice features:
\bea
       \widetilde{F G} &=&\tilde{F} \tilde{G} \cr
       \widetilde{  \left\{ F(t),G(t) \right\}  } &=&
\left\{\tilde{F}(t),\tilde{G}(t)\right\}  \cr
      \left\{ \tilde{F}(t),\phi(t) \right\}  &=& 0  \cr
      \left\{ \tilde{F}(t),\chi(t) \right\}  &=& 0 ~.
\eea
We now define an extended Hamiltonian $H$
and an extended BRST charge $\Omega$ as
\bea
    H(t) &=&  \tilde{H}_o(t) \cr
    \Omega(t)  &=&  i^{\epsilon_A} \chi^i_A(t) {\cal C}^A_i(t) +
\tilde{\Omega}_o(t) ~.
\eea
(No integration over time in the last formula.)
Note that with this choice of Hamiltonian $H$ the shift operation
$\sim$ and total time derivation commute.
It is easy to check that
\bea
    \left\{\Omega(t),\Omega(t) \right\} &=&0 \cr
    \left\{H(t),\Omega(t) \right\}     &=& 0 ~,
\eea
and that the constraints $(\chi^i_A,\tilde{G}_{\alpha})$ form a first class
algebra.
Therefore, the Schwinger-Dyson shift symmetry is fully integrated in the
enlarged phase space.
Note that the only genuine collective shift field is $\varphi^A_{i=1}$.
The new BRST transformation rules with
\mbox{$\delta =\left\{\cdot,\Omega\right\}$} are:
\beq
\begin{array}{rclcrcl}
    \delta \phi^A &=&  i^{\epsilon_A}  {\cal C}^A_1&~~~~~~~~~~~~~&
               \delta {\cal P}^i_A &=& (-i)^{\epsilon_A} \chi^i_A\\
    \delta \varphi^A_1
&=&  i^{\epsilon_A}  {\cal C}^A_1- i^{\epsilon_A}  \tilde{\cal R}^A && \\
     \delta \varphi^A_2 &=&  i^{\epsilon_A}  {\cal C}^A_2 && \delta \chi^i_A
&=&  0  \\
 &&&& \delta \pi_A &=&  \tilde{\cal R}_A\\
    \delta {\cal C}^A_i  &=& 0 && \delta \pi^1_A &=& - \tilde{\cal R}_A \\
 &&&&\delta \pi^2_A &=& 0 ~.
    \end{array}
 \eeq
The extended effective action $S_{\rm eff}$ takes the form:
\beq
   S_{\rm eff} =  \Pi_{\cal A} \dot{ \Phi}^{\cal A}  + \int dt \left(- H
+ \delta \left(\psi_s  +  \psi_{\rm gf} \right)\right) ~,
\eeq
where it is convenient to extract a certain 'shift' fermion $\psi_s$
from the total gauge fermion
\mbox{$\psi_s+\psi_{\rm gf}$}.
\bea
   \psi_s(t)&\equiv& - (-i)^{\epsilon_A} \left( {\cal P}^i_A(t) \dot{
\varphi}^A_i(t)
+ g^j_i {\cal P}^i_A(t)  \varphi^A_j(t) \right) \cr
    g^j_i  &\equiv& g \left| \epsilon^{ij}  \right|
                     \ = \  g \left[ \begin{array}{cc} 0 & 1\\ 1 & 0
\end{array}  \right] ~.
\label{gij}
 \eea
$g$ is a constant of dimension $({\rm time})^{-1}$.
By a straightforward calculation, the extended effective action
$S_{\rm eff}$ can be rewritten in a more useful manner
\beq
S_{\rm eff} = \hat{S} +S_{\delta} +\int dt \ \delta \psi_{\rm gf}  ~.
\eeq
Here we have defined an action $\hat{S}$
\beq
      \hat{S}\equiv \pi_A \dot {\tilde{\phi}}^A -  \int dt \ \tilde{H}_o
                     + {\cal P}^1_A  \tilde{\dot{ {\cal R} }}^A
               + \phi^*_A \tilde{{\cal R}}^A~,
\label{shat}
\eeq
and an auxiliary 'delta function term'
\mbox{$S_{\delta} = S^1_{\delta} +  S^2_{\delta}$}:
\bea
  - S_{\delta}&\equiv& g^j_i \chi^i_A  \varphi^A_j
+ g^j_i {\cal P}^i_A  {\cal C}^A_j  \ = \  - S^1_{\delta} -  S^2_{\delta} \cr
 - S^1_{\delta} &\equiv& g \left( \chi^1_A  \varphi^A_2
     + {\cal P}^1_A  {\cal C}^A_2 \right)
  \ = \ (-i)^{\epsilon_A}  g \delta( {\cal P}^1_A  \varphi^A_2 ) \cr
 - S^2_{\delta} &\equiv& g  \chi^2_A  \varphi^A_1 +\phi^*_A {\cal C}^A_1 ~.
\label{sdelta}
\eea
We have identified antifields $\phi^*_A$ :
\beq
   \phi^*_A \equiv g {\cal P}^2_A ~,
\eeq
as the generators of the field BRST symmetry ${\cal R}^A$ in the
(extended) quantum action $\hat{S}$.

If we now consider an arbitrary function $G=G(\phi,\pi)$
and gauge fixing fermion $ \psi_{\rm gf}=\psi_{\rm gf}(\phi,\pi)$
of the original phase space variables  $(\phi^A,\pi_A)$,
the corresponding Green function in the extended phase space
equals the original Green function
\beq
<G> \ = { { \int \! {\cal D} \Phi {\cal D} \Pi \ G \
e^{{i \over \hbar} \left(  \hat{S} +S_{\delta} +\int dt \ \delta
\tilde{\psi}_{\rm gf}  \right) } }
  \over { \int \! {\cal D} \Phi {\cal D} \Pi \
e^{{i \over \hbar} \left(  \hat{S} +S_{\delta} +\int dt \ \delta
\tilde{\psi}_{\rm gf}  \right) } } }
  = { { \int  \! {\cal D} \phi {\cal D} \pi \ G \
e^{{i \over \hbar}  \left( \pi_A \dot{ \phi}^A
+ \int dt \left(- H_o + \delta_o \psi_{\rm gf} \right)  \right) } }
\over  { \int  \! {\cal D} \phi {\cal D} \pi \
e^{{i \over \hbar}  \left( \pi_A \dot{ \phi}^A
+ \int dt \left(- H_o + \delta_o \psi_{\rm gf} \right)  \right) } }  } \ = \
<G>_o~.
\label{ggo}
\eeq
Therefore {\em the original theory is included in the extended theory}.
This justifies the raison d'\^{e}tre of the whole 'shift' construction.
The proof of  (\ref{ggo}) goes as follows:
First of all, it is convenient
to change the integration variables
in path integrals over the extended
phase space from \mbox{$(\Phi,\Pi)=(\cdots,\pi,\cdots,\pi^i,\cdots)$}
to \mbox{$(\cdots,\pi,\cdots,\chi^i,\cdots)$}.
(The dots indicate unchanged variables.)
The Jacobian of this change of variables is $1$, due to (\ref{chipi}).
Now integration over $\chi^2$, $\varphi_2$, ${\cal C}_2$ and  ${\cal C}_1$,
produces delta functions in
$\varphi_1$, $\chi^1$, ${\cal P}^1$ and  \mbox{$\phi^* \equiv g {\cal P}^2$}
respectively, due to the $S_{\delta}$ term. This proves equation (\ref{ggo}).

The BRST symmetry in the extended phase space is called
Schwinger-Dyson BRST symmetry, because the Schwinger-Dyson equations
\beq
  \int \! {\cal D} \phi \  \left( {\hbar \over i} { {\delta^r G} \over
{\delta\phi^A}}
    + G(\phi) { {\delta^r S_{\rm cl}} \over {\delta\phi^A} }  \right)
e^{{i \over \hbar} \left(  S_{\rm cl} + \delta_{\rm cl}\psi_{\rm gf} (\phi)
\right) }  =  0
\label{sdeq}
\eeq
are fulfilled whenever
\mbox{$ {\hbar \over i} { {\delta^r G} \over {\delta\phi^A}}
    + G(\phi) { {\delta^r S_{\rm cl}} \over {\delta\phi^A} } $}
is BRST closed,
if the boundary condition (\ref{classicallimit}) is satisfied.
The proof is composed  of several steps.
First, we show (in two different ways) that
 \mbox{$ <  {\hbar \over i} { {\delta^r G} \over {\delta\phi^A}}
    + G(\phi) { {\delta^r \hat{S}} \over {\delta\phi^A} }  >  \ = \  0$}.
\bea
 0&=& -i^{\epsilon_A} < \delta ( G(\phi) \phi^*_A) >
   \ = \  <  {  (-i)^{\epsilon_A} {\delta^r G} \over {\delta\phi^B}}
   i^{\epsilon_B}  {\cal C}^B_1 \phi^*_A - G g \chi^2_A >  \cr
  &=& {1 \over {\cal Z} } \int \! {\cal D} \Phi {\cal D} \Pi \
  e^{{i \over \hbar} \left(  \hat{S} +\delta\psi_{\rm gf} \right) }   \left(
 i^{\epsilon_A} i^{\epsilon_B}  { {\delta^r G} \over {\delta\phi^B}} {\cal
C}^B_1
      {\hbar \over i} { {\delta^l} \over {\delta{\cal C}^A_1}}
+ G {\hbar \over i}  { {\delta^r } \over {\delta\varphi^A_1}}  \right)
   e^{{i \over \hbar} S_{\delta}  }  \cr
  &=& {1 \over {\cal Z} } \int \! {\cal D} \Phi {\cal D} \Pi \
  e^{{i \over \hbar} \delta\psi_{\rm gf}  }   \left(
  {\hbar \over i} { {\delta^r G} \over {\delta\phi^B}}  \delta^B_A
+ G\left( {\hbar \over i}  { {\delta^r } \over {\delta\varphi^A_1}}
 -  { {\delta^r \hat{S}} \over {\delta\varphi^A_1} }  \right) \right)
    e^{{i \over \hbar} \left(  \hat{S} +S_{\delta}  \right) }  \cr
   &=&   <  {\hbar \over i} { {\delta^r G} \over {\delta\phi^A}}
    + G(\phi) { {\delta^r \hat{S}} \over {\delta\phi^A} }  >
 \ = \  {1 \over {\cal Z} } \int \! {\cal D} \Phi {\cal D} \Pi \
   {\hbar \over i} { {\delta^r } \over {\delta\phi^A}}
   G  e^{{i \over \hbar} \left(  \hat{S} +S_{\delta} + \delta\psi_{\rm gf}
\right) }    \ = \   0   ~.
\label{sdward}
\eea
In the above\footnote{A comment on notation :
If there is no function appearing in the 'numerator' of a differential
operator,
the differential operators operate on every function appearing to the right.
If there is a function the differential operator only operates on this.}
manipulations (\ref{sdward})
we have assumed that the gauge fermion term
\mbox{$\delta\psi_{\rm gf} $} is independent of $\phi$, $\varphi_1$ and  ${\cal
C}_1$,
and we have performed partial integrations in $\varphi_1$ and  ${\cal C}_1$.
Now it is clear, from arguments similar to (\ref{ggo}), that
\beq
   <  {\hbar \over i} { {\delta^r G} \over {\delta\phi^A}}
    + G(\phi) { {\delta^r \hat{S}} \over {\delta\phi^A} }  >
  \ = \  <  {\hbar \over i} { {\delta^r G} \over {\delta\phi^A}}
    + G(\phi) { {\delta^r \left( \pi_A \dot{ \phi}^A
  - \int dt \ H_o \right)} \over {\delta\phi^A} }  > _o ~.
\eeq
Finally, one integrates out the original momenta $\pi_A$
to obtain the Schwinger-Dyson equations (\ref{sdeq}),
by applying boundary condition (\ref{classicallimit}).
$S_{\rm cl}$ does not depend on the original gauge ghost so one needs to
chose an appropriate gauge fixing term \mbox{$\delta_{\rm cl} \psi_{\rm gf} $}
in order for the gauge ghost integrations in the
Schwinger-Dyson equations (\ref{sdeq}) to become meaningful.
This is allowed because it does not change the correlators.
Here,  \mbox{$\delta_{\rm cl}  $} denotes the original BRST symmetry
in which momenta are integrated out.

\section{An Extended Master Equation in the Standard BRST Case}
\label{stdqme}

Let us now prove a quantum master equation in the extended phase space.
The principle is to impose Ward identities \mbox{$< \delta G >=0$}
for every function \mbox{$G=G(\Phi,\Pi)$} generated by the
Schwinger-Dyson BRST symmetry in the extended phase space.
Consider a BRST variation of an arbitrary function $G=G(\Phi,\Pi)$:
\beq
 \delta G = { {\delta^r G} \over {\delta\phi^A}}  i^{\epsilon_A}  {\cal C}^A_1
   + \left.{ {\delta^r G} \over {\delta\pi_A}}\right|_{\chi} \tilde{{\cal R}}_A
   +  { {\delta^r G} \over {\delta\varphi^A_1}}  i^{\epsilon_A}  \left(  {\cal
C}^A_1
            -  { {\delta^l \hat{S}} \over {\delta\phi^*_A} } \right)
    +  { {\delta^r G} \over {\delta\varphi^A_2}}  i^{\epsilon_A}  {\cal C}^A_2
   +  { {\delta^r G} \over {\delta{\cal P}^i_A}}  (-i)^{\epsilon_A} \chi^i_A
{}~.
\eeq
We have made use of the fact that $\phi^*_A$ only appears linearly
in the action $\hat{S}$ as the generator for  $\tilde{{\cal R}}^A$.
Evaluating the BRST variation inside a path integral
with gauge fixing term $ \psi_{\rm gf}=0$ and using Ward identities,
we find, after performing partial integration in the $G$-derivatives:
\bea
 0&=&- {\cal Z} <\delta G>  \cr
   &=& \int \! {\cal D} \Phi {\cal D} \Pi \ G  \left(
  (-i)^{\epsilon_A} { {\delta^l} \over {\delta\phi^A}} {\cal C}^A_1
    + \left. { {\delta^r} \over {\delta\pi_A}}\right|_{\chi} \tilde{{\cal R}}_A
 \right. \cr
   && \left.   +  (-i)^{\epsilon_A}  { {\delta^l} \over {\delta\varphi^A_1}}
         \left( {\cal C}^A_1 - { {\delta^l \hat{S}} \over {\delta\phi^*_A}}
\right)
+  (-i)^{\epsilon_A}  { {\delta^l} \over {\delta\varphi^A_2}}  {\cal C}^A_2
   -  i^{\epsilon_A} { {\delta^l} \over {\delta{\cal P}^i_A}}  \chi^i_A \right)
   e^{{i \over \hbar} \left(  \hat{S} +S_{\delta}  \right) }  \cr
   &=&  \int \! {\cal D} \Phi {\cal D} \Pi \
   e^{{i \over \hbar} S_{\delta}  } G \left(
  (-i)^{\epsilon_A} { {\delta^l} \over {\delta\phi^A}} {\cal C}^A_1
    + \left. { {\delta^r} \over {\delta\pi_A}}\right|_{\chi} \tilde{{\cal
R}}_A\right. \cr
  && \left.   + \left( -  (-i)^{\epsilon_A} { {\delta^l} \over {\delta\phi^A}}
   - {i^{\epsilon_A+1} g \over \hbar} \chi^2_A   \right)
      \left( {\cal C}^A_1 - { {\delta^l \hat{S}} \over {\delta\phi^*_A}}
\right) \right. \cr
   && \left.  - {i^{\epsilon_A+1} g \over \hbar} \chi^1_A  {\cal C}^A_2
   +i^{\epsilon_A} \chi^i_A  \left( {i g^j_i \over \hbar}  {\cal C}^A_j
    - { {\delta^l} \over {\delta{\cal P}^i_A} } \right) \right)
   e^{{i \over \hbar}   \hat{S} }  \cr
   &=&  \int \! {\cal D} \Phi {\cal D} \Pi \
   e^{{i \over \hbar} S_{\delta}  } G \left( {\hbar \over i}
  (-i)^{\epsilon_A} { {\delta^l } \over {\delta\phi^A} } { {\delta^l}  \over
{\delta\phi^*_A}}
   + \left. { {\delta^r} \over {\delta\pi_A}}\right|_{\chi}  \tilde{{\cal R}}_A
    -i^{\epsilon_A}\chi^1_A   { {\delta^l} \over {\delta{\cal P}^1_A} }
\right)
   e^{{i \over \hbar}   \hat{S} } ~.
\eea
Applying the fundamental lemma in calculus of variations,
we get a quantum master equation in the extended phase space:
\beq
  \left( {\hbar \over i}\Delta
   + \left. { {\delta^r} \over {\delta\pi_A}}\right|_{\chi}  \tilde{{\cal R}}_A
    -i^{\epsilon_A}\chi^1_A   { {\delta^l} \over {\delta{\cal P}^1_A} }
\right)
   e^{{i \over \hbar}   \hat{S} } = 0 ~.
\label{qmee}
\eeq
We have introduced the odd Laplacian
\beq
\Delta \equiv
  (-i)^{\epsilon_A} { {\delta^l } \over {\delta\phi^A} } { {\delta^l}  \over
{\delta\phi^*_A} }   ~.
\eeq
Alternatively, one could derive the quantum master equation (\ref{qmee})
from Batalin-Fradkin-Vilkovisky Theorem,
which states that the partition function ${\cal Z}_{\psi_{\rm gf}}$
is independent of the gauge fixing fermion $\psi_{\rm gf}$.
Choosing an arbitrary infinitesimal\footnote{An infintesimal
variation is denoted with a 'hard $d$'.} variation
$G=d\psi_{\rm gf}$ around $ \psi_{\rm gf}=0$,
the equation
\beq
0= {\cal Z}_{d\psi_{\rm gf}} -  {\cal Z}_0
= \int \! {\cal D} \Phi {\cal D} \Pi \
 e^{{i \over \hbar} \left(  \hat{S} +S_{\delta}  \right) } \  \delta \
d\psi_{\rm gf} \
=  \ <\delta G> ~,
\eeq
would lead to  the same proof and exactly the same conclusions (\ref{qmee}).

It is easy to formally extract the BV Lagrangian theory.
First, use (\ref{sdelta}) to manipulate
the exponent $\hat{S}$ to $\hat{S}+S^1_{\delta} $
in the master equation (\ref{qmee}):
\beq
  \left( {\hbar \over i}\Delta
   + \left. { {\delta^r} \over {\delta\pi_A}}\right|_{\chi}  \tilde{{\cal R}}_A
    -i^{\epsilon_A} \chi^1_A \left(g{\cal C}^A_2
     +  { {\delta^l} \over {\delta{\cal P}^1_A} }  \right) \right)
   e^{{i \over \hbar}  \left( \hat{S} +S^1_{\delta}  \right) }  = 0~.
\label{qmeel}
\eeq
Now, let $\hat{W}$ be defined as
\beq
   e^{ {i \over \hbar}   \hat{W}}
\equiv \int \! {\cal D} {\cal P}^1  {\cal D} {\cal C}_2 {\cal D} \chi^1  {\cal
D} \varphi_2   \
   e^{{i \over \hbar}  \left( \hat{S} +S^1_{\delta}  \right) } ~.
\label{what}
\eeq
Defining the Lagrangian quantum action
\mbox{$W=W\left[\phi-\varphi_1,\phi^* \right]$}:
\beq
   e^{ {i \over \hbar}   W}
\equiv \int \! {\cal D} \pi   \   e^{{i \over \hbar}   \hat{W} }
  \ =  \  \int \! {\cal D} {\cal P}^1  {\cal D} {\cal C}_2
     {\cal D} \chi^1  {\cal D} \varphi_2 {\cal D} \pi   \
   e^{{i \over \hbar}  \left( \hat{S} +S^1_{\delta}  \right) } ~,
\eeq
and integrating over $\pi$, $\varphi_2$, $ \chi^1$,
${\cal C}_2$ and ${\cal P}^1$ in (\ref{qmeel}),
one formally regains the Lagrangian quantum master equation:
\beq
   \Delta    e^{{i \over \hbar}  W}  = 0~.
\label{lqme}
\eeq
Performing the integrations in (\ref{what})
by means of (\ref{shat},\ref{sdelta})
\beq
   \hat{W} = \pi_A \dot {\tilde{\phi}}^A -  \int dt \  \left. \tilde{H}_o
\right|_{\chi=0}
               + \phi^*_A  \left. \tilde{{\cal R}}^A \right|_{\chi=0} ~,
\eeq
one obtains  that  the Lagrangian quantum action becomes
\beq
   e^{ {i \over \hbar}   W}
\equiv \int \! {\cal D} \pi   \   e^{{i \over \hbar}  \left(
 \pi_A \dot {\tilde{\phi}}^A -  \int dt \  \left. \tilde{H}_o  \right|_{\chi=0}
               + \phi^*_A  \left. \tilde{{\cal R}}^A \right|_{\chi=0}
\right)} ~.
\label{lqa}
\eeq
Note that the action $W$  is not necessarily linear in the antifields $\phi^*$
due to the $\pi$  integration (if ${\cal R}^A$ depend on the $\pi$'s).
Imposing the boundary condition (\ref{classicallimit}) yields
\beq
   W\left[\phi-\varphi_1,\phi^*=0 \right] \ = \  S_{\rm
cl}\left[\phi-\varphi_1\right]~.
\eeq

\section{$Sp(2)$ Symmetric Case}
\label{sp2brst}

Let us now turn to the $Sp(2)$ symmetric case.
The shift construction is almost the same although we have to extend
it with twice as many shift fields.
We focus on the new features of the $Sp(2)$ symmetric case,
and leave out the details that are carried over unchanged
from the standard BRST case.
Consider a Hamiltonian system with fields $\phi^A$ and momenta $\pi_A$,
Hamiltonian $H_o$ and {\em two} BRST charges $\Omega^a_o$, $a=1,2$
with corresponding (right) BRST transformations
\mbox{$\delta^a_o =\left\{\cdot,\Omega^a_o\right\}$}.
$a=1$ denotes the standard BRST and $a=2$ is the anti BRST.
The Hamiltonian $H_o$ and the charge $\Omega^a_o$ fulfill
\bea
    \left\{\Omega^a_o(t),\Omega^b_o(t) \right\} &=&0 \cr
    \left\{H_o(t),\Omega^a_o(t) \right\}     &=&0 \cr
    {\partial \over \partial t}\Omega^a_o &=&0~.
\eea
We assume that the BRST transformation rules of the phase space variables
are organized as\footnote{The sign convention of the
$\epsilon$-tensor is $\epsilon^{12}=\epsilon_{21}=+1$.}
\bea
    \delta^a_o \phi^A &=&  i^{\epsilon_A} {\cal R}^{Aa} (\phi,\pi)  \cr
    {1 \over 2}\epsilon_{ab} \delta^a_o\delta^b_o \phi^A &=& i {\cal R}^A
(\phi,\pi)  \cr
    \delta^a_o \pi_A &=&  {\cal R}^a_A (\phi,\pi) ~.
\eea
${\cal R}^A$ is often referred to as the commutator
of the BRST and anti BRST transformation.
A Hamiltonian counterpart to a master equation
is fulfilled in the original theory
\beq
 \left(  i^{\epsilon_A}  { {\delta^r} \over {\delta\phi^A}}  {\cal R}^{Aa}
  +{ {\delta^r} \over {\delta\pi_A}} {\cal R}^a_A  \right)
   e^{{i \over \hbar}   \left(  \pi_A \dot {\phi}^A -  \int dt \  H_o \right)}
= 0 ~.
\label{hme2}
\eeq
We now impose the Schwinger-Dyson BRST symmetry
and extend the phase space with shift fields $\varphi^A_i$, $\pi^A_i$,
shift momenta $\pi^i_A$, $\varphi^i_A$
shift ghost/antighost fields  ${\cal C}^{Aa}_i$,
and shift ghost/antighost momenta ${\cal P}^i_{Aa}$.
Altogether,  we have a ninefold increase in the number of fields:
\beq
\begin{array}{lrcl}
    {\rm Fields} &\Phi^{\cal A}
    &=& \left\{ \phi^A, \varphi^A_i ,\pi^A_i, {\cal C}^{Aa}_i \right\} \\
    {\rm Momenta} &\Pi_{\cal A}
   &=& \left\{ \pi_A, \pi^i_A, \varphi^i_A, {\cal P}^i_{Aa} \right\} ~.
\end{array}
\eeq
Note that $\pi^A_i$ are {\em fields}, and $\varphi^i_A$ {\em momenta}.
The Grassmann parity, ghost number
and new ghost number \cite{bltha2} are given by
\beq
\begin{array}{rcccccccccccl}
    \epsilon(\pi^A_i)&=& \epsilon(\varphi^i_A)&=&
     \epsilon(\pi^i_A)&=&\epsilon(\varphi^A_i)&=&\epsilon(\pi_A)
     &=&\epsilon(\phi^A)&\equiv&\epsilon_A \\
 &&    \epsilon({\cal P}^i_{Aa})&=&\epsilon({\cal C}^{Aa}_i)&=&\epsilon_A+1 \\
    \\
 &&    \gh(\pi^A_i)&=&\gh(\varphi^A_i)&=&\gh(\phi^A)&\equiv&\gh_A \\
  &&   \gh(\varphi^i_A)&=&\gh(\pi^i_A)&=&\gh(\pi_A)&=&-\gh_A \\
   &&  \gh({\cal C}^{Aa}_i)&=&\gh_A - (-1)^a &=&-\gh({\cal P}^i_{Aa})\\
    \\
 &&    \ngh(\varphi^A_i)&=&\ngh(\phi^A)&\equiv&\ngh_A \\
 &&    \ngh(\pi^i_A)&=&\ngh(\pi_A)&=&-\ngh_A \\
 &&   \ngh(\pi^A_i)&=&\ngh_A+2&=& -\ngh(\varphi^i_A) \\
 &&   \ngh({\cal C}^{Aa}_i)&=&\ngh_A +1&=&-\ngh({\cal P}^i_{Aa}) ~.
\end{array}
\eeq
The shift constraints $\chi^i_A$ are
\bea
     \chi^1_A &=&  \pi_A + \pi^1_A  \cr
     \chi^2_A  &=&  \pi^2_A \cr
     \chi^A_1  &=&  \pi^A_1 \cr
     \chi^A_2  &=&  \pi^A_2~.
\eea
The shift operation $\sim$ is unchanged
compared to the standard BRST case,
although it now operates on a larger space.
\beq
F(\phi,\pi,\varphi_1,\pi^1,\varphi_2,\pi^2,\cdots)
\stackrel{\sim}{\longrightarrow}
\tilde{F}\equiv F(\phi-\varphi_1,\pi-\chi^1,0,0,\cdots)
=F(\phi-\varphi_1,-\pi^1,0,0,\cdots)~.
\eeq
We now define an extended Hamiltonian $H$
and  extended BRST charges $\Omega^a$ as
\bea
    H(t) &=&  \tilde{H}_o(t) \cr
    \Omega^a(t)  &=& i^{\epsilon_A} \chi^i_A(t) {\cal C}^{Aa}_i(t)
    +(-i)^{\epsilon_A+1} \epsilon^{ab} \chi^A_i(t) {\cal P}^i_{Ab}(t)
+\tilde{\Omega}_o(t)~.
\eea
(No integration over time in the last formula.)
It is easy to check that
\bea
    \left\{\Omega^a(t),\Omega^b(t) \right\} &=&0 \cr
    \left\{H(t),\Omega^a(t) \right\}     &=& 0 ~,
\eea
and that the constraints $(\chi^i_A, \chi^A_i,\tilde{G}_{\alpha})$ form a first
class algebra.
The BRST transformation rules \mbox{$\delta^a =\left\{\cdot,\Omega^a \right\}$}
are:
\beq
\begin{array}{rclcrclcrcl}
     \delta^a \phi^A &=& i^{\epsilon_A} {\cal C}^{Aa}_1 &~~~~~~~~~~~~~&
                \delta^a \varphi^i_A &=& i^{\epsilon_A+1} \epsilon^{ab}{\cal
P}^i_{Ab}
                 &~~~~~~~~~~~~~&    \delta^a \pi^A_i &=&  0 ~.\\
    \delta^a \varphi^A_1 &=& i^{\epsilon_A} {\cal C}^{Aa}_1
        -i^{\epsilon_A} \tilde{\cal R}^{Aa}  && \\
    \delta^a \varphi^A_2 &=& i^{\epsilon_A} {\cal C}^{Aa}_2
              && \delta^a {\cal P}^i_{Ab} &=&  \delta ^a_b (-i)^{\epsilon_A}
\chi^i_A   \\
\\
     \delta^a {\cal C}^{Ab}_i  &=& (-i)^{\epsilon_A+1}\epsilon^{ab} \chi^A_i
           &&\delta^a \chi^i_A &=&  0  \\
&&&& \delta^a \pi_A &=&  \tilde{\cal R}^a_A \\
\delta^a \chi^A_i &=&  0&&\delta^a \pi^1_A &=& - \tilde{\cal R}^a_A \\
&&&&  \delta^a \pi^2_A &=& 0
    \end{array}
 \eeq
The action $S_{\rm eff}$ takes the form
\bea
   S_{\rm eff} &=&  \Pi_{\cal A} \dot{ \Phi}^{\cal A}  + \int dt \left(- H
 +    {1 \over 2}\epsilon_{ab} \delta^a \delta^b \left( \chi_s+\chi_{\rm gf}
\right) \right) \cr
 &=& \hat{S} +S_{\delta}
 +\int dt \  {1 \over 2}\epsilon_{ab} \delta^a \delta^b \chi_{\rm gf}  ~.
\label{seff2}
\eea
We have  extracted a 'shift' boson $\chi_s$
from the total gauge boson
$\chi_s+\chi_{\rm gf}$:
\beq
   \chi_s(t) \equiv i\varphi^i_A(t) \dot{ \varphi}^A_i(t) + i g^j_i
\varphi^i_A(t)  \varphi^A_j(t) ~.
\label{shiftboson}
\eeq
$g^j_i $ is defined as in equation (\ref{gij}).
In the second line of (\ref{seff2}), we have introduced an action
$\hat{S}$ and a 'delta function' part \mbox{$S_{\delta} = S^1_{\delta} +
S^2_{\delta}$}:
\bea
      \hat{S}&\equiv&\pi_A \dot {\tilde{\phi}}^A -  \int dt \ \tilde{H}_o
                      + \varphi^1_A  \tilde{\dot{ {\cal R} }}^A
               +\bar{\phi}_A \tilde{{\cal R}}^A
                    + {\cal P}^1_{Aa}  \tilde{\dot{ {\cal R} }}^{Aa}
               + \phi^*_{Aa} \tilde{{\cal R}}^{Aa}  \cr
 - S_{\delta}&\equiv& g^j_i \chi^i_A  \varphi^A_j
+g^j_i \varphi^i_A  \chi^A_j
+g^j_i {\cal P}^i_{Aa}  {\cal C}^{Aa}_j   \ = \  - S^1_{\delta} -  S^2_{\delta}
\cr
- S^1_{\delta}&\equiv&
 g \left( \chi^1_A  \varphi^A_2+\varphi^1_A  \chi^A_2
   + {\cal P}^1_{Aa}  {\cal C}^{Aa}_2 \right)
   \ = \  -  {ig \over 2}\epsilon_{ab} \delta^a \delta^b
             \left(  \varphi^1_A  \varphi^A_2 \right) \cr
-  S^2_{\delta} &\equiv& g \left( \chi^2_A  \varphi^A_1
   + \varphi^2_A  \chi^A_1  \right)
+\phi^*_{Aa}  {\cal C}^{Aa}_1 ~.
\label{2sdelta}
\eea
Here we have identified
%\footnote{The
%definition of $\bar{\phi}_A$ and ${\cal R}^A$ differs with a sign compared to
%$\bar{\phi}_A$ and $Y^A$ respectively of reference \cite{bltla2}.}
$\bar{\phi}_A$ and antifields $\phi^*_{Aa}$
\bea
   \bar{\phi}_A&\equiv& g \varphi^2_A \cr
  \phi^*_{Aa} &\equiv& g {\cal P}^2_{Aa}
  \eea
as the generators of ${\cal R}^A$ and ${\cal R}^{Aa}$ respectively.

If we now take an arbitrary function $G=G(\phi,\pi)$
and a gauge fixing boson $ \chi_{\rm gf}=\chi_{\rm gf}(\phi,\pi)$
of the original phase space variables  $(\phi^A,\pi_A)$,
the corresponding Green function in the extended phase space
equals the original Green function,
because integration over
$\chi^2$, $\varphi_2$, $\chi_2$,  $\chi_1$,  ${\cal C}_2$ and  ${\cal C}_1$,
produces delta functions in
$\varphi_1$, $\chi^1$,$\varphi^1$, \mbox{$\bar{\phi} \equiv g \varphi^2$},
${\cal P}^1$
and  \mbox{$\phi^* \equiv g {\cal P}^2$}
respectively, due to the $S_{\delta}$ term.
\bea
<G> &=&{ { \int \! {\cal D} \Phi {\cal D} \Pi \ G \
e^{{i \over \hbar} \left(  \hat{S} +S_{\delta}
+\int dt \ {1 \over 2} \epsilon_{ab} \delta^a \delta^b \tilde{\chi}_{\rm gf}
\right) }  }
\over  { \int \! {\cal D} \Phi {\cal D} \Pi \
e^{{i \over \hbar} \left(  \hat{S} +S_{\delta}
+\int dt \ {1 \over 2} \epsilon_{ab} \delta^a \delta^b \tilde{\chi}_{\rm gf}
\right) }  } } \cr
  &=& { { \int  \! {\cal D} \phi {\cal D} \pi \ G \
e^{{i \over \hbar}  \left( \pi_A \dot{ \phi}^A
+ \int dt \left(- H_o +{1 \over 2} \epsilon_{ab} \delta^a_o \delta^b_o
 \chi_{\rm gf} \right)  \right) } }
\over   { \int  \! {\cal D} \phi {\cal D} \pi  \
e^{{i \over \hbar}  \left( \pi_A \dot{ \phi}^A
+ \int dt \left(- H_o +{1 \over 2} \epsilon_{ab} \delta^a_o \delta^b_o
 \chi_{\rm gf} \right)  \right) } }  } \ =  \   <G>_o~.
\eea
Again, we see that all physical relevant quantities of
the original Hamiltonian theory can be reproduced
in the extended Hamiltonian theory.

\section{An Extended Master Equation in the $Sp(2)$-symmetric Case}
\label{sp2qme}

Consider a BRST variation of an arbitrary function $G=G(\Phi,\Pi)$:
\bea
 \delta^a G &=& { {\delta^r G} \over {\delta\phi^A}} i^{\epsilon_A} {\cal
C}^{Aa}_1
   + \left.{ {\delta^r G} \over {\delta\pi_A}}\right|_{\chi} \tilde{{\cal
R}}^a_A
   +  { {\delta^r G} \over {\delta\varphi^A_1}} i^{\epsilon_A}\left(  {\cal
C}^{Aa}_1
            -  { {\delta^l \hat{S}} \over {\delta\phi^*_{Aa}} } \right)
    +  { {\delta^r G} \over {\delta\varphi^A_2}} i^{\epsilon_A} {\cal C}^{Aa}_2
 \cr
  && +  { {\delta^r G} \over {\delta{\cal C}^{Ab}_i}}
(-i)^{\epsilon_A+1}\epsilon^{ab} \chi^A_i
+ { {\delta^r G} \over {\delta\varphi^i_A}} i^{\epsilon_A+1} \epsilon^{ab}{\cal
P}^i_{Ab}
   +  { {\delta^r G} \over {\delta{\cal P}^i_{Aa}}}  (-i)^{\epsilon_A} \chi^i_A
 ~.
\eea
We have made use of the fact that $\phi^*_{Aa}$ only appears linearly
in the action $\hat{S}$  as the generator for  $\tilde{{\cal R}}^{Aa}$.
Evaluating the BRST variation inside a path integral with
gauge fixing term $ \chi_{\rm gf}=0$ and using Ward identities,
we find, after performing partial integration in the $G$-derivatives:
\bea
 0&=&-{\cal Z} <\delta^a G>  \cr
   &=& \int \! {\cal D} \Phi {\cal D} \Pi \ G  \left(
  (-i)^{\epsilon_A} { {\delta^l} \over {\delta\phi^A}} {\cal C}^{Aa}_1
    + \left. { {\delta^r} \over {\delta\pi_A}}\right|_{\chi}  \tilde{{\cal
R}}^a_A  \right. \cr
   && \left.   +  (-i)^{\epsilon_A}  { {\delta^l} \over {\delta\varphi^A_1}}
         \left( {\cal C}^{Aa}_1 - { {\delta^l \hat{S}} \over
{\delta\phi^*_{Aa}}} \right)
+  (-i)^{\epsilon_A}  { {\delta^l} \over {\delta\varphi^A_2}}  {\cal C}^{Aa}_2
\right. \cr
  && \left.  +i^{\epsilon_A+1} \epsilon^{ab} { {\delta^l} \over {\delta{\cal
C}^{Ab}_i}}  \chi^A_i
- (-i)^{\epsilon_A+1} { {\delta^l} \over {\delta\varphi^i_A}} \epsilon^{ab}
{\cal P}^i_{Ab}
   - i^{\epsilon_A} { {\delta^l} \over {\delta{\cal P}^i_{Aa}}}  \chi^i_A
\right)
   e^{{i \over \hbar} \left(  \hat{S} +S_{\delta}  \right) }  \cr
   &=&  \int \! {\cal D} \Phi {\cal D} \Pi \
   e^{{i \over \hbar} S_{\delta}  } G \left(
  (-i)^{\epsilon_A} { {\delta^l} \over {\delta\phi^A}} {\cal C}^{Aa}_1
    + \left. { {\delta^r} \over {\delta\pi_A}}\right|_{\chi}  \tilde{{\cal
R}}^a_A \right. \cr
  && \left.   + \left( -  (-i)^{\epsilon_A} { {\delta^l} \over {\delta\phi^A}}
   - {i^{\epsilon_A+1} g \over \hbar} \chi^2_A  \right)
      \left( {\cal C}^{Aa}_1 - { {\delta^l \hat{S}} \over {\delta\phi^*_{Aa}}}
\right)
- {i^{\epsilon_A+1}  g \over \hbar} \chi^1_A  {\cal C}^{Aa}_2  \right. \cr
  && \left.  - { (-i)^{\epsilon_A} g^i_j \over \hbar} \epsilon^{ab} {\cal
P}^j_{Ab} \chi^A_i
+ (-i)^{\epsilon_A+1} \left( {i g^j_i \over \hbar} \chi^A_j
- { {\delta^l} \over {\delta\varphi^i_A}}  \right) \epsilon^{ab} {\cal
P}^i_{Ab}
   + i^{\epsilon_A}\chi^i_A  \left( {i g^j_i \over \hbar}  {\cal C}^{Aa}_j
    - { {\delta^l} \over {\delta{\cal P}^i_{Aa}} } \right) \right)
   e^{{i \over \hbar}   \hat{S} }  \cr
   &=&  \int \! {\cal D} \Phi {\cal D} \Pi \
   e^{{i \over \hbar} S_{\delta}  } G \left( {\hbar \over i}
  (-i)^{\epsilon_A} { {\delta^l } \over {\delta\phi^A} } { {\delta^l}  \over
{\delta\phi^*_{Aa}}}
-  (-i)^{\epsilon_A+1}\epsilon^{ab} {\cal P}^i_{Ab} { {\delta^l} \over
{\delta\varphi^i_A}} \right. \cr
  && \left.   + \left. { {\delta^r} \over {\delta\pi_A}}\right|_{\chi}
\tilde{{\cal R}}^a_A
    - i^{\epsilon_A}\chi^1_A   { {\delta^l} \over {\delta{\cal P}^1_{Aa}} }
\right)
   e^{{i \over \hbar}   \hat{S} } ~.
\eea
Applying the fundamental lemma in calculus of variation,
we get a quantum master equation in the extended phase space:
\beq
  \left( {\hbar \over i}\Delta^a + V^a
   + \left. { {\delta^r} \over {\delta\pi_A}}\right|_{\chi} \tilde{{\cal
R}}^a_A
-  (-i)^{\epsilon_A+1}\epsilon^{ab} {\cal P}^1_{Ab} { {\delta^l} \over
{\delta\varphi^1_A}}
   - i^{\epsilon_A}\chi^1_A   { {\delta^l} \over {\delta{\cal P}^1_{Aa}} }
\right)
   e^{{i \over \hbar}   \hat{S} } = 0 ~.
\label{2qmee}
\eeq
We have introduced the odd Laplacian $\Delta^a$ and vector field $V^a$:
\bea
\Delta^a &\equiv &
  (-i)^{\epsilon_A} { {\delta^l } \over {\delta\phi^A} } { {\delta^l}  \over
{\delta\phi^*_{Aa}} }  \cr
V^a &\equiv& - (-i)^{\epsilon_A+1}\epsilon^{ab} \phi^*_{Ab}
     { {\delta^l} \over {\delta \bar{\phi}_A}}
 \ =  \  - (-i)^{\epsilon_A+1}\epsilon^{ab} {\cal P}^2_{Ab}
     { {\delta^l} \over {\delta\varphi^2_A}} ~.
\eea
Let us derive the Lagrangian theory of Batalin,  Lavrov and Tyutin
\cite{bltla2}.
First, use (\ref{2sdelta}) to manipulate the exponent $\hat{S}$ to
$\hat{S}+S^1_{\delta} $
in the master equation (\ref{2qmee}):
\beq
   \left( {\hbar \over i}\Delta^a + V^a
   + \left. { {\delta^r} \over {\delta\pi_A}}\right|_{\chi} \tilde{{\cal
R}}^a_A
- (-i)^{\epsilon_A+1}\epsilon^{ab} {\cal P}^1_{Ab}  \left( g \chi^A_2
       + { {\delta^l} \over {\delta\varphi^1_A}}  \right)
   - i^{\epsilon_A}\chi^1_A  \left(g{\cal C}^A_2
    +   { {\delta^l} \over {\delta{\cal P}^1_{Aa}} }  \right)  \right)
   e^{{i \over \hbar}  \left( \hat{S} +S^1_{\delta}  \right) } = 0 ~.
\label{2qmeel}
\eeq
Now, let $\hat{W}$ be defined as
\beq
   e^{ {i \over \hbar}   \hat{W}}
\equiv \int \!  {\cal D} \chi_2 {\cal D} \varphi^1
      {\cal D} {\cal P}^1  {\cal D} {\cal C}_2 {\cal D} \chi^1  {\cal D}
\varphi_2   \
   e^{{i \over \hbar}  \left( \hat{S} +S^1_{\delta}  \right) } ~.
\label{2what}
\eeq
The Lagrangian quantum action
$W=W\left[\phi-\varphi_1,\bar{\phi},\phi^* \right]$ is defined as
\beq
   e^{ {i \over \hbar}   W}
\equiv \int \! {\cal D} \pi   \   e^{{i \over \hbar}   \hat{W} }
  \ =  \  \int \!  {\cal D} \chi_2 {\cal D} \varphi^1
        {\cal D} {\cal P}^1  {\cal D} {\cal C}_2
     {\cal D} \chi^1  {\cal D} \varphi_2 {\cal D} \pi   \
   e^{{i \over \hbar}  \left( \hat{S} +S^1_{\delta}  \right) } ~.
\eeq
The Lagrangian quantum master equation is formally reproduced by integrating
over
$\pi$, $\varphi_2$, $ \chi^1$, ${\cal C}_2$, ${\cal P}^1$, $\varphi^1$ and
$\chi_2$
in (\ref{2qmeel}):%
\beq
  \left({\hbar \over i} \Delta^a + V^a\right)
         e^{{i \over \hbar}   W } = 0 ~.
\label{lqme2}
\eeq
The integrations in (\ref{2what}) is easily done with the help of
(\ref{2sdelta})
\beq
   \hat{W} = \pi_A \dot {\tilde{\phi}}^A -  \int dt \  \left. \tilde{H}_o
\right|_{\chi=0}
               + \bar{\phi}_A  \left. \tilde{{\cal R}}^A \right|_{\chi=0}
               + \phi^*_{Aa}  \left. \tilde{{\cal R}}^{Aa} \right|_{\chi=0} ~,
\eeq
Therefore, the Lagrangian quantum action is
\beq
   e^{ {i \over \hbar}   W}
\equiv \int \! {\cal D} \pi   \   e^{{i \over \hbar}  \left(
\pi_A \dot {\tilde{\phi}}^A -  \int dt \  \left. \tilde{H}_o  \right|_{\chi=0}
               + \bar{\phi}_A  \left. \tilde{{\cal R}}^A \right|_{\chi=0}
               + \phi^*_{Aa}  \left. \tilde{{\cal R}}^{Aa} \right|_{\chi=0}
\right)} ~.
\eeq
The classical action $S_{\rm cl}$ is regained for $\phi^*_{Aa}=0=\bar{\phi}_A$
if one  impose the boundary condition (\ref{classicallimit}).

\section{Conclusion}
We have shown in great detail how to introduce
the Lagrangian antifields and quantum action in a generic Hamiltonian theory
assuming no second class constraint and using canonical flat (Darboux)
coordinates.
This has been done both in the standard BRST case and in case of $Sp(2)$
symmetry
by introducing collective shift fields.
As a spin-off we have derived a new quantum master equation
(\ref{qmee},\ref{2qmee}) in an extended phase space.
It is pleasant that the Lagrangian and Hamiltonian pictures
merge together in this extended phase space.

It is amazing how smooth the $Sp(2)$ discussion extends the standard BRST case.
This is a priori not at all obvious.
The major difference is that in order to have a manifest
$Sp(2)$ BRST exact expression,
it has to be inside {\em both} the BRST transformations
$\delta^1$ {\em and} $\delta^2$.
The magic is the shift boson (\ref{shiftboson}) quadratic in the fundamental
fields.
Each boson term leads to $4$ terms in the action,
so if one wants to modify the action
by $1$ term one has to alter with $3$ other terms as well,
because modifications should be $Sp(2)$ BRST exact .

The discussion of boundary conditions has some loose ends.
A more complete treatment would require a more explicit analysis of the
original theory.
One should bear in mind that the classical action is often not unique.
In this context, it would be desirable to have a procedure enforcing manifest
Lorentz-covariance for the Lagrangian quantum action.
Another problem is how to avoid the assumption of flat coordinates, i.e.
to have a covariant formulation.

\vspace{0.5cm}
\noindent

{\sc Acknowledgment:} The author would like to thank
P.H. Damgaard for many fruitful discussions.
This work  was supported by NorFA grant no. 95.30.074-O.
\newpage
%\vspace{1cm}

\end{document}